\RequirePackage{lineno}
\documentclass[twocolumn,showpacs,preprintnumbers,amsmath,amssymb,prl,superscriptaddress]{revtex4}
\usepackage{graphicx}
\usepackage{dcolumn}
\usepackage{bm}
\usepackage{hyperref}
\modulolinenumbers[1]
\newenvironment{color}[3]
{
}{
}

\newcommand{\grey}[1]     {}

\newcommand{\pT}{$p_{\rm{T}}$}

\newcommand{\sNN}{$\sqrt {{s_{\rm NN}}}$}

\newcommand{\energy}{$\langle \epsilon  \rangle$}
\newcommand{\temp}{$\langle T \rangle$}
\newcommand{\Teff}{$T_{\rm eff}$}
\newcommand{\Teffave}{$\langle T_{\rm eff} \rangle$}
\newcommand{\yphi}{$y$--$\phi$} 
\begin{document}
\title{Maps of the Little Bangs Through \\ 
Energy Density and Temperature Fluctuations}

\author{Sumit Basu$^1$, Rupa Chatterjee$^1$, Basanta K. Nandi$^2$ and Tapan K. Nayak}
\vspace{0.2cm}
\affiliation{Variable Energy Cyclotron Centre, 1/AF Bidhan Nagar,
  Kolkata - 700064, India \\
 $^2$Indian Institute of Technology Bombay, Mumbai - 400076, India}
\medskip

\date{\today}

\begin{abstract} 

In this letter we propose for the first time to map the heavy-ion
collisions at ultra-relativistic energies, similar to the maps of the cosmic microwave background
radiation, using fluctuations of energy density and
temperature in small phase space bins.
We study the evolution of fluctuations at each stage of the collision
using an event-by-event
hydrodynamic framework. 
We demonstrate the
feasibility of making fluctuation maps from experimental data 
and its usefulness in extracting 
considerable information regarding the early stages of
the collision and its evolution.

\end{abstract}


\maketitle


Observation of the cosmic microwave background radiation (CMBR)
by various satellites confirms the Big Bang evolution,
inflation and provide important information regarding the early
Universe and its evolution with excellent accuracy~\cite{cmbr1,cmbr2,cmbr3}. 
The physics of heavy-ion collisions at ultra-relativistic energies,
popularly known as little bangs, has often been 
compared to the Big Bang 
phenomenon of early Universe~\cite{Book,heinz,morphology,paul,ajit}. 
The matter produced at extreme conditions of energy density ($\epsilon$) and
temperature ($T$) in heavy-ion collisions is a Big Bang replica in a tiny scale.
In little bangs, the produced fireball goes through a rapid evolution
from an early state of partonic quark-gluon plasma (QGP) to a hadronic
phase, and finally freezes out within a few tens of fm.  
Heavy-ion experiments are predominantly sensitive to the conditions that
prevail at the later stage of the collision as majority of the
particles are emitted near the freeze-out.
As a result, a direct and
quantitative estimation of the properties of hot and dense matter 
in the early stages and during each stage of the evolution has not yet been possible. 

In this letter, 
we propose to make temperature fluctuation maps, 
and use fluctuation measures to quantitatively probe the early stages of the
heavy-ion collisions.
We demonstrate the making of fluctuation maps 
in bins of rapidity~($y$) and azimuthal angle~($\phi$) from the AMPT event 
generator~\cite{ampt}, which is a proxy for experimental data. 
These maps can be used to make a detailed analysis similar to
those of the CMBR fluctuation 
methods~\cite{heinz,ajit, paul, morphology}. 
We use hydrodynamics to model the evolution of the produced system and 
make maps of $\epsilon$ and $T$ from initial time to freeze-out, 
and to estimate their relative fluctuations. 
By making a correspondence of measured fluctuations with the time
evolution profiles of the fluctuations from
hydrodynamic calculations, 
we show that it is possible to visualize
the thermodynamic conditions of colliding matter that presumably existed
at different stages of evolution.
Furthermore, we take advantage of the large number of particles
produced in the heavy-ion collisions 
to obtain transverse momentum~(\pT) distribution and extract the
temperature from each event. We demonstrate that
event-by-event temperature fluctuations can be used to provide
important thermodynamic parameters for the system created in 
the collisions~\cite{Sto,korus,steph,gavai-gupta,wilk,shuryak,morw}.

\begin{figure*}[tbp]
\centering
\includegraphics[width=0.90\textwidth]{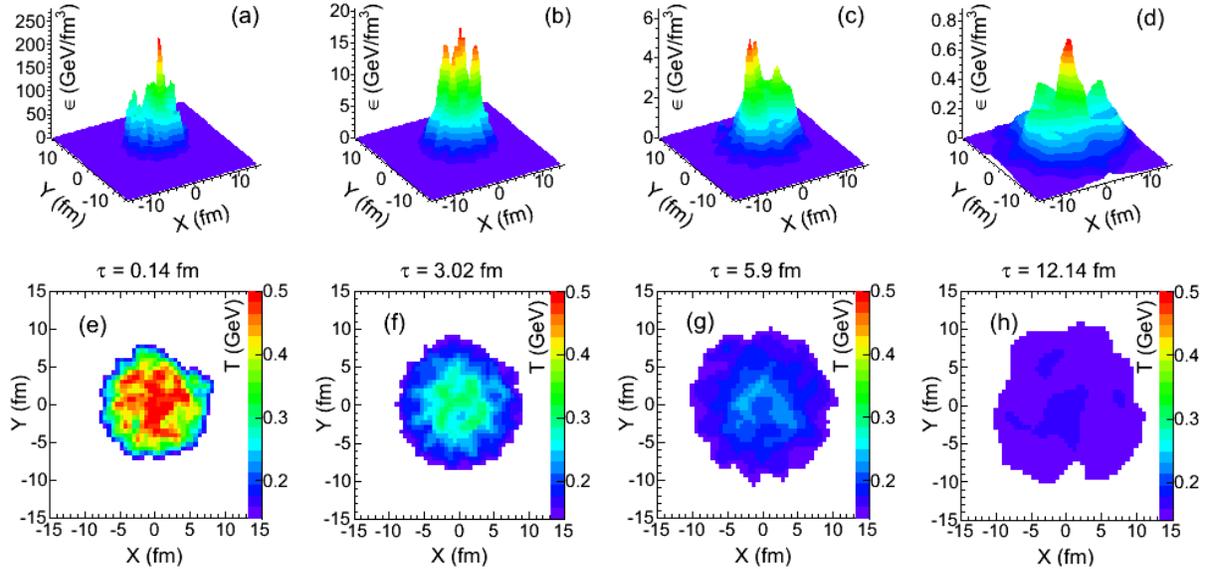} 
\caption
{(Color online). Distributions of
energy density (upper panles) and temperature
(lower panels) in the transverse ($X$-$Y$) plane at four proper times 
($\tau$), obtained from hydrodynamic calculations for one central
Pb-Pb event at \sNN~=2.76~TeV.
}
\label{fig1}
\end{figure*}

Recent experimental data from the Relativistic Heavy Ion Collider (RHIC)
and the Large Hadron Collider (LHC) have confirmed the formation of a
strongly coupled system.
Hydrodynamics has been used extensively and to a large extent 
successfully to explain majority 
of these experimental results~\cite{song}.
A (2+1)-dimensional  event-by-event ideal hydrodynamical
framework~\cite{hannu} 
is used in the present work to model the
space-time evolution of the system produced in most central (0--5\% of
the total cross section)
collisions of lead nuclei at \sNN~=~2.76~TeV at LHC.
A lattice-based equation of state~\cite{eos} is used in this model 
and 170 MeV is considered as the transition temperature from the QGP
to a hadronic phase. 
This model has been successfully used to explain the spectra and
elliptic flow of hadrons at RHIC and LHC energies~\cite{hannu}.
In this Monte Carlo Glauber model, the standard two-parameter
Woods-Saxon nuclear density profile is used to distribute the nucleons
randomly into the colliding nucleons. Two nucleons from different nuclei
are assumed to collide when $d^2 < \sigma_{\rm NN}/\pi$, where $d$ is the
transverse distance between the nuclei and $\sigma_{\rm NN}$ is the
inelastic nucleon-nucleon cross section. For LHC energies, we take $\sigma_{\rm NN} = 64$ mb
and the initial formation time of the plasma $\tau_0=$
0.14~fm~\cite{phe,hannu, chre1}.
A wounded nucleon (WN) profile is considered where the initial
entropy density is distributed around the WN using a 2-dimensional
Gaussian distribution function,
\begin{equation}
 s(X,Y) = \frac{K}{2 \pi \sigma^2} \sum_{i=1}^{\ N_{\rm WN}} 
\exp \Big( -\frac{(X-X_i)^2+(Y-Y_i)^2}{2 \sigma^2} \Big).
\label{eq:eps}
\end{equation}
Here $X_i,Y_i$ are the transverse coordinates of the $i^{\rm th}$ nucleon and $K$ is an
overall normalization constant. The size of the density fluctuations
is  determined by the  free parameter $\sigma$, which is taken to be
0.4~fm~\cite{hannu}. The temperature at freeze-out 
is taken as 160~MeV, which reproduces the measured \pT~spectra of
charged pions at LHC.

Figure~\ref{fig1} shows the distributions of $\epsilon$ and $T$
for a single event 
in $X$-$Y$ bins (each bin is of size 0.6~fm$\times$0.6~fm) 
at four different values of proper time ($\tau$).
The upper panels (a-d) show a three dimensional view of 
$\epsilon$, whereas the lower panels (e-h) show the~$T$ variations in the
transverse plane.
At early times, sharp and pronounced peaks in
$\epsilon$ and hotspots in $T$ are observed.
These bin-to-bin fluctuations in $\epsilon$ and
$T$ indicate that the system formed immediately after collision is
inhomogeneous in phase space and quite violent. As time elapses, the system
cools, expands, and the bin-to-bin variations in $\epsilon$ and $T$
smoothens out tending towards a homogenous system at freeze-out. 
\begin{figure}[b]
\centering
\includegraphics[width=0.52\textwidth]{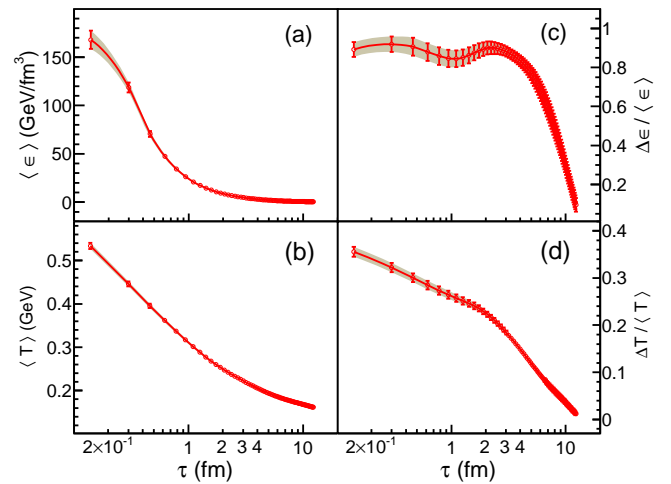} 
\caption{(Color online). Temporal 
evolution of (a) average energy density, (b) average temperature,
(c) fluctuations in energy density, and (d) fluctuations in temperature, 
for central Pb-Pb 
collisions at \sNN~=2.76~TeV, obtained from hydrodynamic calculations.
The shaded regions represent the extent of event-by-event variations.
}
\label{fig2}
\end{figure}

\begin{figure}[tbp]
\centering
\includegraphics[width=0.4\textwidth]{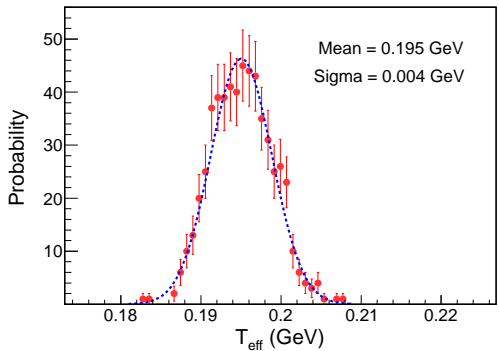} 
\caption{(Color online). 
Bin-to-bin distribution of 
\Teff~of pions for 16$\times$36 bins in \yphi~for central
Pb-Pb collisions 
at \sNN~=~2.76~TeV using the AMPT model. 
}
\label{fig3}
\end{figure}

\begin{figure}[tbp]
\centering
\includegraphics[width=0.49\textwidth]{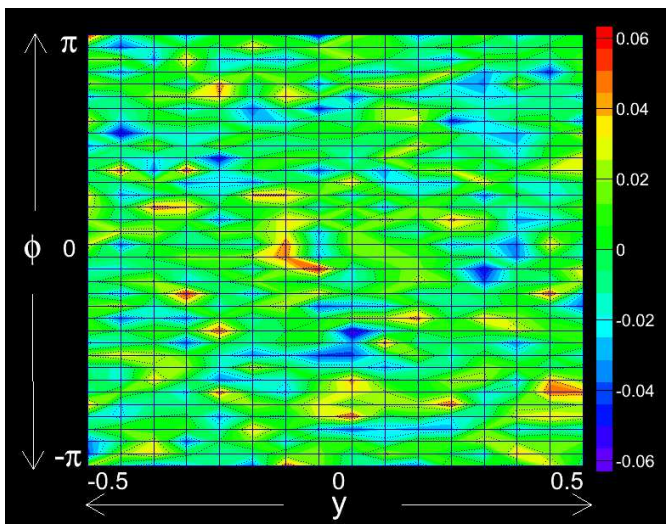} 
\caption{(Color online). 
Temperature fluctuation map
in \yphi~phase space 
for central (0--5\%) Pb-Pb collisions at
\sNN~=~2.76~TeV using the AMPT model.
The right hand color palettes show the magnitude of fluctuations.
}
\label{fig4}
\end{figure}

Observations from Fig.~\ref{fig1} can be quantified by studying the mean 
energy density (\energy), mean temperature (\temp) over all the
bins, and the
bin-to-bin fluctuations of $\epsilon$ and {$T$} at each $\tau$. 
Figure~\ref{fig2} presents the time evolution of \energy~and \temp, 
and their fluctuations. The $x$-axes are plotted in logarithmic scale
for zooming in on the early times. 
The event-by-event variations of these quantities are represented by
the shaded regions, taken from about 500 events.
The \energy~and \temp~decrease as time elapses.
The value of \energy~falls sharply from $\sim$168~GeV/fm$^3$ at
$\tau=0.14$~fm to a value of $\sim$20~GeV/fm$^3$ at $\tau=1$~fm, and
then falls slowly till freeze-out. 
The initial energy density valus are close to the results from 
the ALICE collaboration, $\epsilon$$\tau$$\sim$16~GeV/fm$^2$~\cite{alice_energy}.
On the other hand, the fall of \temp~with $\tau$ is smooth, which goes down
from $\sim$530~MeV at
$\tau=0.14$~fm to $\sim$300~MeV at $\tau=1$~fm.  
At the freeze-out, \temp~is close to 160~MeV.

The bin-to-bin fluctuations in $\epsilon$ and $T$  have been quantified by 
$\Delta \epsilon/\langle \epsilon \rangle$ and $\Delta T/\langle T \rangle$, 
where $\Delta \epsilon$ and  $\Delta T$ are the root mean
square (RMS) deviations.
Time evolutions of fluctuations in $\epsilon$ and $T$ are presented in the 
right panels of Fig.~\ref{fig2}. 
Extremely large fluctuations in energy density of $\sim$90\%
are observed at early times, confirming the violent nature of the
collision. At the same time, the fluctuations in $T$ are smaller ($\sim$35\%). 
Interestingly, although \energy~decreases quite fast, 
the fluctuation in $\epsilon$ remains almost constant up to
$\tau$$\sim$2.5~fm, then falls rapidly. Around the same value of
$\tau$, the fluctuation in $T$ shows a kink, where the change in
fluctuation increases. 
There may be a characteristic change in the behaviour of the system
at this $\tau$ during the hydrodynamic evolution.
A close analysis of the time evolution of different phases indicates
that these changes in nature of the fluctuations happen at a time when
the hadronic phase starts to dominate over the QGP phase.

Fluctuation measurements in heavy-ion experiments are possible only at
freeze-out. The connection from
the freeze-out to the early stages of colision can be made by
comparing experimental data with theoretical calculations at freeze-out.
For this comparison, the available phase space 
of experimental data is divided into bins of $y$-$\phi$. 
For each bin, the  \pT~spectrum of identified
particles can be constructed, from which one can extract the effective
temperature (\Teff). 
The bin-to-bin fluctuation of \Teff~are used to construct fluctuation
maps and attempt to make a connection to the hydrodynamic calculations.
The applicability of this method is demonstrated below
using the AMPT event generator.

We employ the string melting (SM) mode of the AMPT model~\cite{ampt}
to mimic the experimental conditions for Pb-Pb collisions at \sNN~=~2.76~TeV.
This mode includes a fully partonic QGP phase that hadronizes through
quark coalescence, and has been shown to reproduce the experimental
data at LHC energies~\cite{subrata,cmko1,cmko2,dronika}.  The
parameter set-B as discussed in~\cite{cmko1,cmko2} 
is used for the event generation. 
We choose events with 0--5\% centrality window to ensure that the 
event-to-event variation in the number of participating nucleons is minimal.

The phase space around the central rapidity ($-0.5 \le y \le 0.5$) and
full azimuth ($-\pi \le \phi \le \pi$)
is divided into a number of bins in \yphi. We present
results of the calculation with a 
grid of 16$\times$36 bins, 
where the bin sizes are well within the
detector resolutions for the present experiments at RHIC and LHC.
The \pT~spectrum of charged pions are constructed 
for each \yphi~bin by combining a large number of events. 
For each of the bins, the \pT~spectrum is fitted by a Maxwell-Boltzmann
function within $0.5 \le p_{\rm T} \le 1.0$~GeV
to obtain \Teff.
This \pT~range is chosen to exclude very
low \pT~pions coming from resonance decays and high \pT~pions affected
by mini-jets~\cite{alice}. 
\Teff~has contributions from two components, a thermal part and a second part
which depends on the collective transverse velocity~($\langle
\beta_{\rm T} \rangle$) of the system.
Assuming that the second component is similar for all the events within a
narrow centrality event class, the fluctuation in \Teff~can be considered to be a good
representation of the fluctuation in kinetic temperature.
Figure~\ref{fig3} shows the distribution of \Teff~for all the bins,
fitted with a Gaussian distribution. The mean and the sigma of the distribution
are 195~MeV and 3.8~MeV
respectively, which represents a fluctuation of 2\% over all the bins.

We construct the fluctuation map in \yphi~bins, where the
temperature fluctuation in each bin is calculated in terms of
$\Delta T_{\rm eff}/\langle T_{\rm eff}\rangle$, where
$\Delta T_{\rm eff} = (\langle T_{\rm eff} \rangle - T_{\rm eff})$.
Here \Teff~is the effective temperature of a particular bin and
\Teffave~is the average value over all bins.
Figure~\ref{fig4} shows the fluctuation map  for all the
\yphi~bins and the amount of fluctuations from the mean value is
represented by different colors.
The color palettes are smeared in the profile histogram for a better
representative view.

The map gives a quantitative view of the temperature fluctuations in
the available phase space. 
It clearly shows several hot (red) as well as cold (blue) spots,
with average (green) zones throughout the phase space.
These spots may have their origin from the extreme regions of phase space,
which existed during the early stages of the reaction. This may
indicate that the observed fluctuations  are remnants of the initial
energy density fluctuations and are not washed out until the freeze-out stage. 
The amount of these
fluctuations are similar to those from hydrodynamic calculations at
$\tau\sim$12~fm.
Thus, within the present theoretical framework, 
we can make a correspondence to the early stage of the collision
through the hydro calculations.
Furthermore, these fluctuation maps can form the basis of 
power spectrum analysis~\cite{ajit, paul, morphology}.

\begin{figure}[tbp]
\centering
\includegraphics[width=0.4\textwidth]{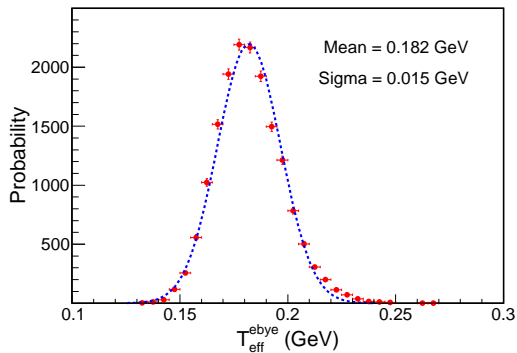} 
\caption{(Color online). 
Event-by-event distribution of $T^{\rm ebye}_{\rm eff}$ for pions within
central rapidity and full azimuth for central Pb-Pb collisions
at \sNN~=~2.76~TeV using AMPT model. 
}
\label{fig5}
\end{figure}

In CMBR studies of the Big Bang, there is only one event,
whereas in the case of heavy-ion collision experiments, one has access to
very large number of events. This can be used as an advantage in order
to gain access to the primordial state. 
Event-by-event fluctuations
are sensitive to the changes in the
state of the matter, and  give additional thermodynamic
information. Event-by-event temperature fluctuations are related to the
heat capacity ($C_{\rm v}$) of the system~\cite{Sto,korus,steph,gavai-gupta,wilk,shuryak,morw}:
\begin{eqnarray}
C_{\rm v}^{-1} = (\Delta T^{\rm ebye}_{\rm eff}/ \langle T^{\rm ebye}_{\rm eff} \rangle)^2,
\end{eqnarray}
where $T^{\rm ebye}_{\rm eff}$ is the effective temperature, obtained
on an event-by-event basis. At LHC energies, central 
Pb-Pb collisions produce a large number of particles in each event, 
which makes the event-by-event analysis of several
observables within the reach of the experiments.
We have constructed \pT~distributions of charged pions 
on an event-by-event basis 
for a large number of AMPT events corresponding to 0-5\%
centrality. The \pT~spectrum for each event is fitted with
Maxwell-Boltzmann function to obtain $T^{\rm ebye}_{\rm eff}$.
The distribution of $T^{\rm ebye}_{\rm eff}$, fitted with a Gaussian distribution, is shown in
Fig.~\ref{fig5}, which gives the mean and sigma as
182~MeV and 15~MeV respectively, which yields $C_{\rm v}=147$
for the system at freeze-out. This method opens up a new avenue for accessing the
thermodynamic parameters.

We discuss the following effects which can affect temperature fluctuations:

{\it Mean \pT}:
Similar fluctuation maps may be constructed from fluctuation of mean
transverse momentum ($\langle p_{\rm T} \rangle$) of charged particle
spectra~\cite{na49}. However, 
$\langle p_{\rm T} \rangle$ may not be a good measure of the
temperature~\cite{steph}. 

{\it Event plane orientation}:
Event plane orientation is necessary for studying
bin-to-bin fluctuations, especially for non-central collisions. For
the present study, AMPT events are event plane oriented.

{\it Flow effect}:
Fourier decomposition of the momentum distribution in transverse plane 
yields a $\phi$--independent, axially symmetric radial flow component and a
$\phi$--dependent part containing the anisotropic flow coefficients. 
For most central collisions within a narrow centrality bin, 
radial flow remains similar for all the events and the anisotropic
flow components do not affect the slope of the \pT~distribution. 

{\it Final state effects}: 
Final state effects, such as resonance decay, and hadronic rescattering
tend to make the \pT~spectra softer, mostly below 0.5~GeV. 
To mitigate this effect, the fit range is chosen to be 0.5~GeV to 1.0~GeV.

{\it Finite multiplicity effect}:
Pb-Pb collisions at LHC energies produce a large number of particles
which are adequate for event-by-event studies. 
The number bins in constructing the map should not be too large in order to avoid
the empty bin effect.

{\it Number of \yphi~bins}: 
The limitation of choosing the
number of bins is constrained by detector resolutions of the various
experiments. 
This is tested the by choosing
several \yphi~bins (16$\times$16, 16$\times$24 and 16$\times$36). 
The bin-to-bin temperature fluctuations remain within 
1.3\% to 2\%. 

{\it Event averaging}:
Temperature fluctuation map involves 
constructing the \pT~spectrum in a 
given \yphi~bin by including
particles from a large number of events. As the final spectrum is event
averaged, the averaging does not affect the determination of slope parameters.

{\it Species dependence}: 
As the particle production mechanism of baryons, mesons and strange
particles are different, species dependence of 
temperature fluctuations may provide extra
information of their freeze-out hyper-surfaces. 
Whether the origin of the 
temperature fluctuations are solely due to initial state
fluctuations or any final state effect,
this will be interesting to study in the species dependence of
temperature fluctuations. 

{\it Viscosity effect}:
Viscosity tends to dilute the fluctuations. 
The SM version of AMPT includes the effect of viscosity 
($\eta/s \sim 0.15$ at $T$=436~MeV~\cite{subrata,dronika}).
Analysis using a viscous hydrodynamic
model is in progress.

In conclusion, we have shown that temperature fluctuation maps,
similar to those in CMBR experiments, offer a novel way of
representing data in heavy-ion collisions.
Experimentally, it is possible to obtain
bin-to-bin fluctuations in temperature
from the transverse momentum distributions in \yphi~bins.
Interestingly, quantitative similar fluctuations are obtained from
hydrodynamic model calculations for most central Pb-Pb collisions at
the LHC. Non-zero fluctuations imply that some of the signals of the
initial state fluctuations remain as imprints at the freeze-out, this
connection has been established within the framework of a hydrodynamic model.
Important information like the speed of sound, specific heat, etc.,
can be extracted from event-by-event temperature
fluctuations.
We emphasize that this novel way of studying temperature fluctuations
will open new avenues of studying heavy-ion collisions and will
be useful in obtaining proper insight into the little bang and QGP matter.


\vspace{0.5 cm}
We would like to thank H. Holopainen for providing us with the event-by-event 
hydrodynamic code. 
We gratefully acknowledge stimulating discussions with
Satyajit Jena, 
Alexander Philipp Kalweit, 
Steffen Bass, 
Dinesh Srivastava, 
Sudhir Raniwala,
Paolo Giubellino,
Jurgen Schukraft, 
Zi-Wei Lin, 
Daniel McDonalds and
Bikash Sinha. 

\end{document}